\documentclass[twoside,twocolumn,bibnotes,tightenlines,aps,floats]{revtex4}
\usepackage{amsmath}
\usepackage{graphicx}
\usepackage{times}
\graphicspath{{../Figures/}{.}}
\begin{document}
\bibliographystyle{revtex}
\date{\today}
 
\title{\bf 
Phase Dynamics of Nearly Stationary Patterns in Activator-Inhibitor Systems
}

\author{Aric Hagberg}
\email{aric@lanl.gov} 
\affiliation{
Center for Nonlinear Studies and T-7,
Theoretical Division, Los Alamos National Laboratory, 
Los Alamos, NM 87545
}
\author{Ehud Meron}
\email{ehud@bgumail.bgu.ac.il}
\affiliation{
The Jacob Blaustein Institute for Desert Research 
and the Physics Department,
Ben-Gurion University, Sede Boker Campus 84990, Israel
}
\author{Thierry Passot}
\affiliation{
Observatoire de la Cote d'Azur, BP 4229, 06304 Nice Cedex 4, France,
and Department of Mathematics, 
University of Arizona, Tucson, AZ 85721
}

\begin{abstract}
The slow dynamics of nearly stationary patterns in a FitzHugh-Nagumo
model are studied using a phase dynamics approach. A Cross-Newell
phase equation describing slow and weak modulations of periodic
stationary solutions is derived. The derivation applies to the
bistable, excitable, and the Turing unstable regimes. 
In the bistable case stability thresholds
are obtained for the Eckhaus and the zigzag instabilities and
for the transition to traveling waves.  Neutral stability curves demonstrate 
the destabilization of stationary planar patterns at low wavenumbers 
to zigzag and traveling modes. Numerical solutions of the model system 
support the theoretical findings.
 
\end{abstract}

\revised{January 26, 2000}


\maketitle

\section{Introduction} 
Studies of stationary patterns in activator-inhibitor systems
have focused primarily on localized structures such as pulses and
spots in excitable and bistable
media~\cite{TyKe:88,KeOs:89,meron:92,KM:94,LeSw:95,MuOs:96,WSBOP:96,SOBP:97},
and periodic patterns near a Turing bifurcation~\cite{CDBD:90,OuSw:91n,AAP:97}.
Localized structures have instabilities to 
traveling patterns, breathing motion,
and transverse deformations~\cite{KeOs:89,OMK:89,HaMe:94a,GMP:96}. 
Periodic patterns have been analyzed 
near the onset of a Turing instability and 
also near the codimension-two
point of a Turing instability and a 
Hopf bifurcation~\cite{RoMe:92,HBP:93,PDDK:93,DLDB:96,OB:98}.  
But very few studies have explored instabilities of {\em periodic}
(nonlocal) stationary patterns in excitable and bistable media, or of
periodic stationary patterns {\em far beyond} the Turing 
instability~\cite{DK:89,Osipov:96}.  
The latter case includes
pattern formation studies on the CIMA chemical 
reaction~\cite{PDDK:93,OuSw:91,DP:94}.

In this paper we study instabilities of stationary periodic patterns
by deriving a Cross-Newell phase equation~\cite{KBBC:82,CN:84,NPBEI:96}. The
derivation is not restricted to the immediate
neighborhood of a Turing instability and applies to periodic
patterns with space-scale separation that arise far from onset or in
excitable and bistable media.  The Cross-Newell equation was
originally derived in the context of fluid dynamics and has recently
been applied in a laser system~\cite{LMN:96}.

We choose to study the FitzHugh-Nagumo (FHN) equations, a canonical
model for activator-inhibitor systems,
\begin{eqnarray}
\label{fhn}
\frac{\partial u}{\partial t}&=&u-u^3-v+\nabla^2 u\,, \\
\frac{\partial v}{\partial t}&=&\epsilon(u-a_1v-a_0)+ \delta\nabla^2 v
\,.  \nonumber
\end{eqnarray}
Here, $u$ is the activator and $v$ the inhibitor.  The parameters
$a_0$ and $a_1$ can be chosen so that the FHN model~(\ref{fhn})
represents an excitable medium, a bistable medium, or a system
with a Turing instability~\cite{HaMe:94a}. All three cases
support stationary periodic solutions for $\delta$ sufficiently
large.

In Section~\ref{phaeqn} we derive a phase equation describing weak
modulations of periodic stripe pattern in the FHN model. In
Section~\ref{stabthre} we evaluate stability thresholds for Eckhaus
and zig-zag instabilities and for a transition from stationary to
traveling patterns. These thresholds suggest a number of spatial or
spatio-temporal behaviors which we test in Section~\ref{simulations}
with numerical solutions of Eqs.~(\ref{fhn}).

\section{The Phase Equation}
\label{phaeqn}

Let $u_0(\theta;k)=u_0(\theta+2\pi;k)$,
$v_0(\theta;k)=v_0(\theta+2\pi;k)$ be a stationary periodic solution
of Eqs.~(\ref{fhn}) with  phase $\theta$ and wavenumber $k$. 
We consider weak spatial modulations of this periodic pattern 
and assume that those modulations have a 
length scale $L$ that is much larger than the wavelength $1/k$.
The ratio of the length scales $\lambda=1/(kL)$ can then be
used as a small parameter to write modulated solutions
as an asymptotic expansion about the periodic
solution
\begin{eqnarray}
u(\theta,{\bf R},T)=u_0(\theta;k)+\lambda u_1(\theta,{\bf R},T) +
\lambda^2 u_2(\theta,{\bf R},T) + ...\nonumber \\ v(\theta,{\bf
R},T)=v_0(\theta;k)+\lambda v_1(\theta,{\bf R},T)+ \lambda^2
v_2(\theta,{\bf R},T) + ...
\label{exp}
\end{eqnarray}
where
${\bf R=\lambda{\bf r}}$ and $T=\lambda^2 t$ are slow space and time
variables.
The phase $\theta$ in Eq.~(\ref{exp}) is an undetermined function of space
and time and $k=|{\bf k}|=|\nabla\theta|$ is the local wavenumber.
Our objective is to derive an equation for the slow phase
\[
\Theta({\bf R},T):=\lambda\theta({\bf r},{\bf R},T)\,.
\]
In terms of this phase the local wavevector is
\[
{\bf k}({\bf R},T)=\nabla_R\Theta\,.
\]
Inserting the expansions~(\ref{exp}) in Eqs.~(\ref{fhn}) we find at order
unity
\begin{subequations}
\label{order_0}
\begin{eqnarray}
u_0-u_0^3-v_0+ k^2 \frac{\partial^2 u_0}{\partial\theta^2}&=&0\,,
\label{order_0_U}\\ 
\epsilon(u_0-a_1v_0-a_0)+ \delta k^2\frac{\partial^2
v_0}{\partial\theta^2} &=&0 \,,
\label{order_0_V}
\end{eqnarray}
\end{subequations}
where $k^2={\bf k}\cdot{\bf k}$. At order $\lambda$ 
\begin{subequations}
\label{order_1}
\begin{eqnarray}
\left(k^2\frac{\partial^2}{\partial\theta^2}+1-3u_0^2\right)u_1 - v_1
&=&{\cal D}\frac{\partial u_0}{\partial\theta}\,, \\ \epsilon u_1 +
\left(\delta k^2 \frac{\partial^2}{\partial\theta^2} - \epsilon
a_1\right)v_1 &=&{\cal D}\frac{\partial v_0}{\partial\theta}\,,
\end{eqnarray}
\end{subequations}
where
\begin{eqnarray}
{\cal D}=\frac{\partial\Theta}{\partial T}-\nabla_R\cdot{\bf k} -2{\bf
k}\cdot\nabla_R\,.
\end{eqnarray}
Projecting the right hand side of (\ref{order_1}) onto
$(\partial_\theta u_0, -\epsilon^{-1}\partial_\theta v_0)$,
the solution of the adjoint problem, produces the phase equation
\[
\tau\frac{\partial\Theta}{\partial T}=-\nabla_R\cdot({\bf k}B)\,,
\]
where
\begin{eqnarray}
\label{tau}
\tau&=&<(\partial_\theta u_0)^2>-\epsilon^{-1}<(\partial_\theta
v_0)^2>\,, \\ B&=&-<(\partial_\theta u_0)^2>+\delta\epsilon^{-1}
<(\partial_\theta v_0)^2>\,.
\label{B}
\end{eqnarray}
In these equations $<(.)>:=\frac{1}{2\pi}\int_0^{2\pi}(.)d\theta$.

The quantities $B$ and $\tau$ contain information about various
instabilities of the periodic stripe pattern. The condition
$\frac{d}{dk}[kB(k)]=0$ implies the onset of an Eckhaus instability
and the condition $B=0$ the onset of a zigzag
instability~\cite{CN:84}.  In Appendix A we show that the condition
$\tau=0$ indicates a transition to traveling waves.

To implement these conditions we need to solve Eqs.~(\ref{order_0})
for the periodic solution $(u_0,v_0)$. For parameter values that
satisfy $\epsilon/\delta:=\mu\ll 1$ an approximate solution can be
computed as shown in Appendix B. Using this solution to calculate
$\tau$ and $B$, as shown in Appendix C, gives the following expressions:
\begin{eqnarray}
\tau&=&\frac{2\sqrt{2}}{3\pi k} - \frac{v_-}{q\pi
k\eta}\beta(\Lambda_-) \gamma(\Lambda_-,\Lambda_+)\,, \nonumber \\
B&=&\frac{2\sqrt{2}}{3\pi k} - \frac{v_-}{q\pi
k\sqrt{\mu}}\beta(\Lambda_-) \gamma(\Lambda_-,\Lambda_+)\,,
\label{tauB}
\end{eqnarray}
\begin{equation}
\Lambda_-+\Lambda_+=\frac{2\pi\sqrt{\mu}}{k} \,,
\label{lpm1}
\end{equation}
\begin{equation}
v_+\beta(\Lambda_+)+v_-\beta(\Lambda_-)=0\,,
\label{lpm2}
\end{equation}
where $\mu=\epsilon/\delta$, $\eta=\sqrt{\epsilon\delta}$, $v_\pm=(\pm
1-a_0)/q^2$, $q^2=a_1+1/2$,
\begin{equation}
\beta(x)=\coth{qx}-{\rm csch}{qx}\,,
\label{beta}
\end{equation} 
and
\begin{eqnarray}
\gamma(\Lambda_-,\Lambda_+)&=&-1+\frac{1}{2}(1+a_0)q\Lambda_-{\rm
csch}{q\Lambda_-}
\nonumber\\ 
&&\mbox{} +\frac{1}{2}(1-a_0)q\Lambda_+{\rm
csch}{q\Lambda_+}\,.
\end{eqnarray}
The quantities $\Lambda_+$ and $\Lambda_-$ denote the widths of
domains with high and low values of $u$ and $v$, respectively. The
width is measured with respect to the spatial coordinate
$z=\frac{\sqrt{\mu}}{k}\theta$ (see Appendix B).  Given $k$,
Eqs.~(\ref{lpm1})
and~(\ref{lpm2}) can be solved for $\Lambda_+(k)$ and
$\Lambda_-(k)$. Using these solutions in Eq.~(\ref{tauB}) graphs of $\tau$
and $kB$ as functions of $k$ can be produced.

\section{Stability Thresholds}
\label{stabthre}
\begin{figure}
\center \includegraphics[width=3.25in]{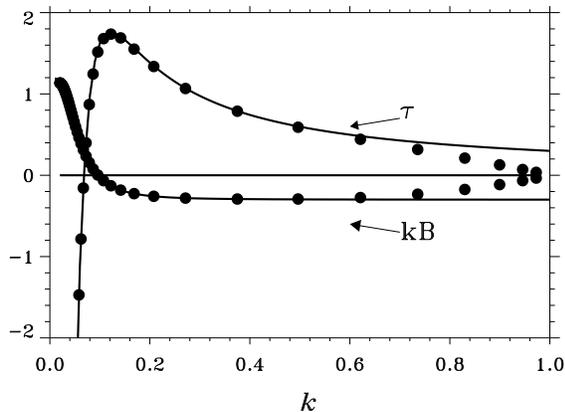} 
\caption{
Typical functions $\tau(k)$ and $kB(k)$ for a bistable medium.  The
curves represent the functions of Eqs.~(\protect\ref{tau_sym_bis}). The
circles are numerically computed solutions using Eqs.~(\ref{tau})
and~(\ref{B}).  The point $kB=0$ indicates the boundary between stable
stationary stripes and zigzag patterns.  At $\tau=0$ the pattern
becomes unstable to traveling waves.  Parameters: $a_1=4$, $a_0=0$,
$\epsilon=0.001$, $\delta=2.0$.  }
\label{fig:tkb-bistable}
\end{figure}

Explicit forms for $\tau(k)$ and $B(k)$ are available in the
symmetric case, $a_0=0$, where
$\Lambda_+=\Lambda_-=\frac{\pi\sqrt{\mu}}{k}$:
\begin{eqnarray}
\label{tau_sym_bis}
\tau(k)=\frac{1}{\pi k\eta_c q^3}\left[1-\frac{\eta_c}{\eta}f(\pi
q\sqrt{\mu}/k)\right]\,, \nonumber \\ B(k)= \frac{1}{\pi k\eta_c
q^3}\left[1-\frac{\eta_c}{\sqrt{\mu}}f(\pi q\sqrt{\mu}/k)\right]\,,
\label{B_sym_bis}
\end{eqnarray}
where $\eta_c=\frac{3}{2\sqrt{2}q^3}$ and
\begin{equation}
f(x)=(1-x~{\rm csch}{x})(\coth{x}-{\rm csch}{x})\,.
\end{equation}
Figure~\ref{fig:tkb-bistable} shows graphs of $\tau(k)$ and $kB(k)$
for a bistable medium obtained with Eqs.~(\ref{tau_sym_bis}) (thick
lines) and with Eqs.~(\ref{tau}) and~(\ref{B}) using numerically
calculated solutions $u_0, v_0$ (circles). A very good agreement is
obtained within the validity range of the analysis, $k\sim {\cal
O}(\sqrt{\mu})\ll 1$. For $k\sim{\cal O}(1)$ the deviations become
large. In particular the minimum of $kB(k)$ which designates the
Eckhaus instability threshold, is not reproduced by the analytical
form~(\ref{B_sym_bis}).

The instability to traveling waves occurs at $\tau=0$ or at
\begin{equation}
\epsilon=\eta_c^2 f^2(\pi q\sqrt{\mu}/k)\delta^{-1}\,.
\label{trav_thres}
\end{equation}
The zigzag instability occurs at $B=0$ or at
\begin{equation}
\epsilon=\eta_c^2 f^2(\pi q\sqrt{\mu}/k)\delta \,.
\label{zigzag_thres}
\end{equation} 
The condition $\frac{d}{dk}(kB)=0$ for the Eckhaus instability becomes
\[
\frac{df}{dx}|_{x=\pi q\sqrt{\mu}/k}=0\,.
\]
Consider first the limit $k\to 0$ in which the periodic pattern
approaches an array of isolated front structures. In this limit $f(\pi
q\sqrt{\mu}/k)\to 1$ and the condition for the onset of traveling
waves becomes $\epsilon=\eta_c^2\delta^{-1}$.  This is precisely the
nonequilibrium Ising-Bloch (NIB) bifurcation point, where a stationary front
loses stability to a pair of counter-propagating fronts.  The
condition for the zigzag instability becomes
$\epsilon=\eta_c^2\delta$. This is the threshold for the transverse
front instability~\cite{HaMe:94c}.

\begin{figure}
{\center 
\includegraphics[width=3.25in]{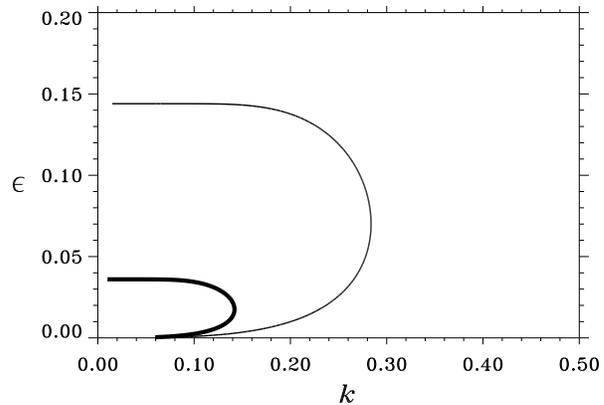}
\includegraphics[width=3.25in]{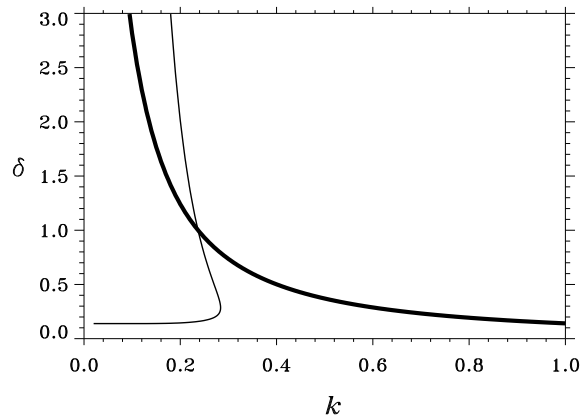} 
}
\caption{
The neutral stability boundaries for the zigzag instability
($B=0$; thick, solid curve) and traveling wave instability 
($\tau=0$; thin, dashed curve).   To the left of the $B=0$
curve planar periodic patterns are unstable to zigzag patterns.
To the left of the $\tau=0$ curve planar periodic patterns are
unstable to traveling waves.
Parameters: $a_1=2$, $a_0=0$, $\epsilon=0.01$, $\delta=2$.
}
\label{fig:neutral}
\end{figure}

The neutral stability curves for a bistable medium corresponding to
Eqs.~(\ref{trav_thres}) and~(\ref{zigzag_thres}) are shown in
Figs.~\ref{fig:neutral}a and ~\ref{fig:neutral}b for fixed $\delta$
and $\epsilon$, respectively. They imply that high wavenumber
stationary planar patterns are stabilized against zigzag and traveling
wave instabilities.  Notice that for $\delta=1$ the neutral stability
curves $\tau=0$ and $B=0$ coincide (see Eqs.~(\ref{trav_thres}) and
(\ref{zigzag_thres}) or Fig.~\ref{fig:neutral}b).  For $\delta>1$,
upon decreasing the wavenumber at constant $\epsilon$, a high
wavenumber pattern is destabilized to a zigzag pattern, whereas for
$\delta<1$ the destabilization is to traveling waves.  Similar neutral
stability curves are found for the nonsymmetric case, $a_0\ne 0$,
for excitable media and for systems (far) beyond the Turing
instability.

%Also shown are neutral stability
%curves (including the Eckhaus curve) obtained from Eqs.~(\ref{tau})
%and~(\ref{B}) using numerically calculated $u_0, v_0$ solutions (thin
%lines). These curves divide the $\epsilon - k$ plane into ??
%regions. ....  Note that the zig-zag instability region refers to
%stationary patterns which might be unstable in some part of this
%region.

\section{Comparisons with Numerical Solutions}
\label{simulations}
\begin{figure}
\center \includegraphics[width=3.0in]{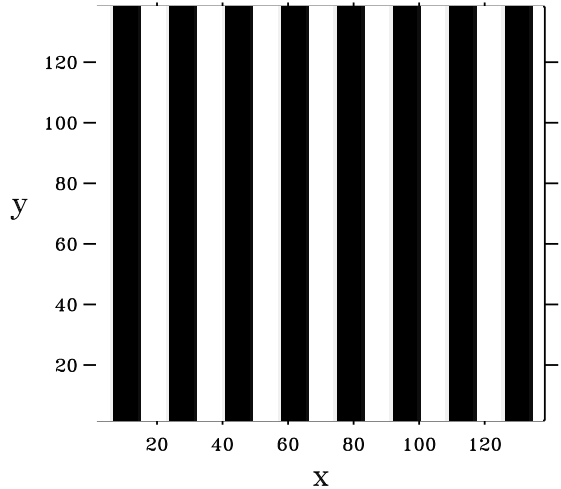}
\includegraphics[width=3.0in]{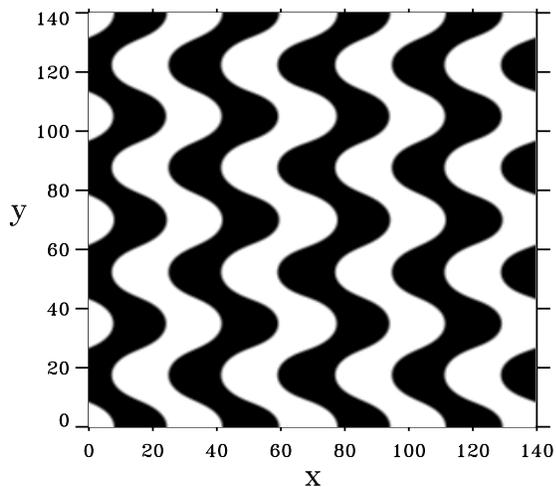}
\caption{
Coexistence of zigzag and planar patterns.
The dark areas indicate regions of $u>0$ and the light
regions $u<0$.
At higher wavenumbers (top) the planar stripe solution is stable.
At low wavenumbers
(bottom) the planar solution is unstable and forms a zigzag pattern.
Parameters: $a1=2$, $a_0=0$, $\epsilon=0.05$, $\delta=2$.  }
\label{fig:zigzag}
\end{figure}

\begin{figure}
\center \includegraphics[width=3.0in]{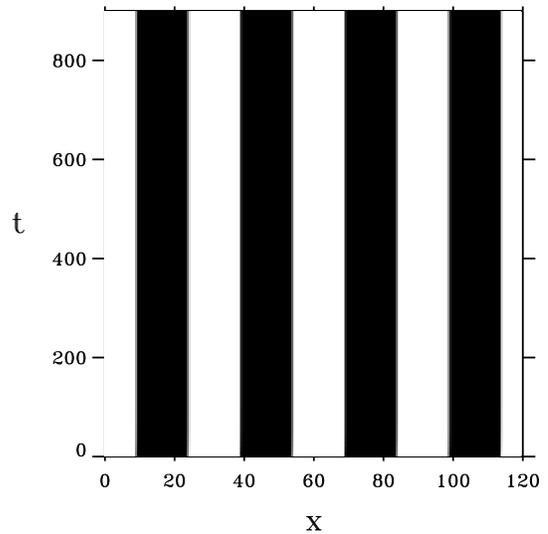}
\includegraphics[width=3.0in]{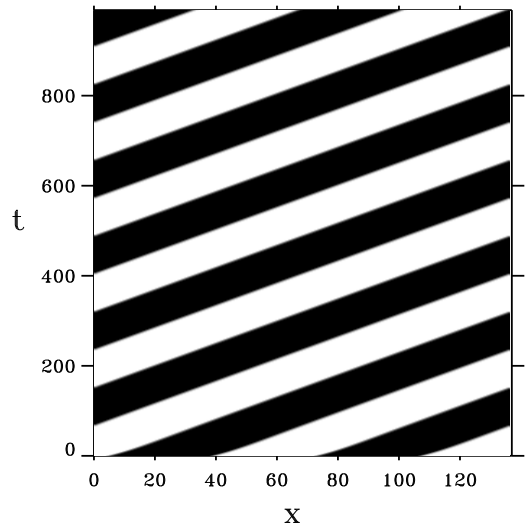}
\caption{
Coexistence of traveling waves and stationary waves.
At high wavenumber (top) the patterns are stationary and at low
wavenumber patterns (bottom) they travel.  Parameters: $a1=2$, $a_0=0$,
$\epsilon=0.03$, $\delta=2$.  }
\label{fig:traveling}
\end{figure}

We have computed numerical solutions of Eqs.~(\ref{fhn})  to
test the stabilization of zigzag and traveling-wave instabilities at
high wavenumbers.  Figure~\ref{fig:zigzag} shows a low wavenumber zigzag
pattern and a high wavenumber planar pattern computed for the same
parameter values.  This behavior is well known in other
contexts~\cite{CrHo:93}.  The zigzag instability is a mechanism by
which the system locally increases the wavenumber.
Figure~\ref{fig:traveling} shows coexistence of a low
wavenumber traveling wave and a high wavenumber stationary pattern.
These numerical results are for a bistable system but similar results
are found for excitable and Turing unstable systems.  
Coexistence of stationary and traveling waves has been found in
experiments on the CIMA reaction~\cite{PDDK:93,DP:94}.
and analyzed using different theoretical
approaches~\cite{DK:89,IO:92,KO:95,Osipov:96}.

\begin{figure}
\center \includegraphics[width=3.0in]{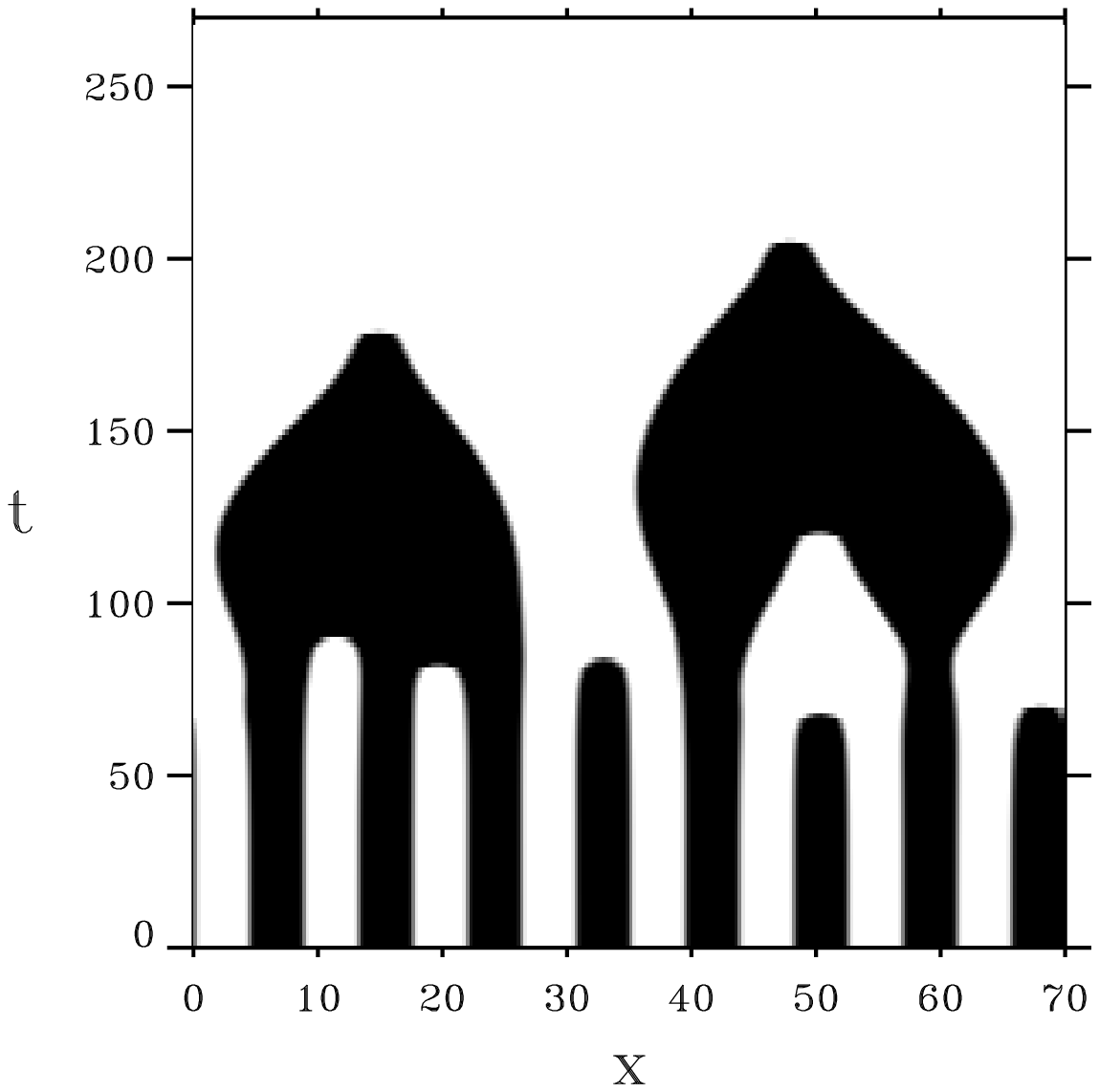}
\includegraphics[width=3.0in]{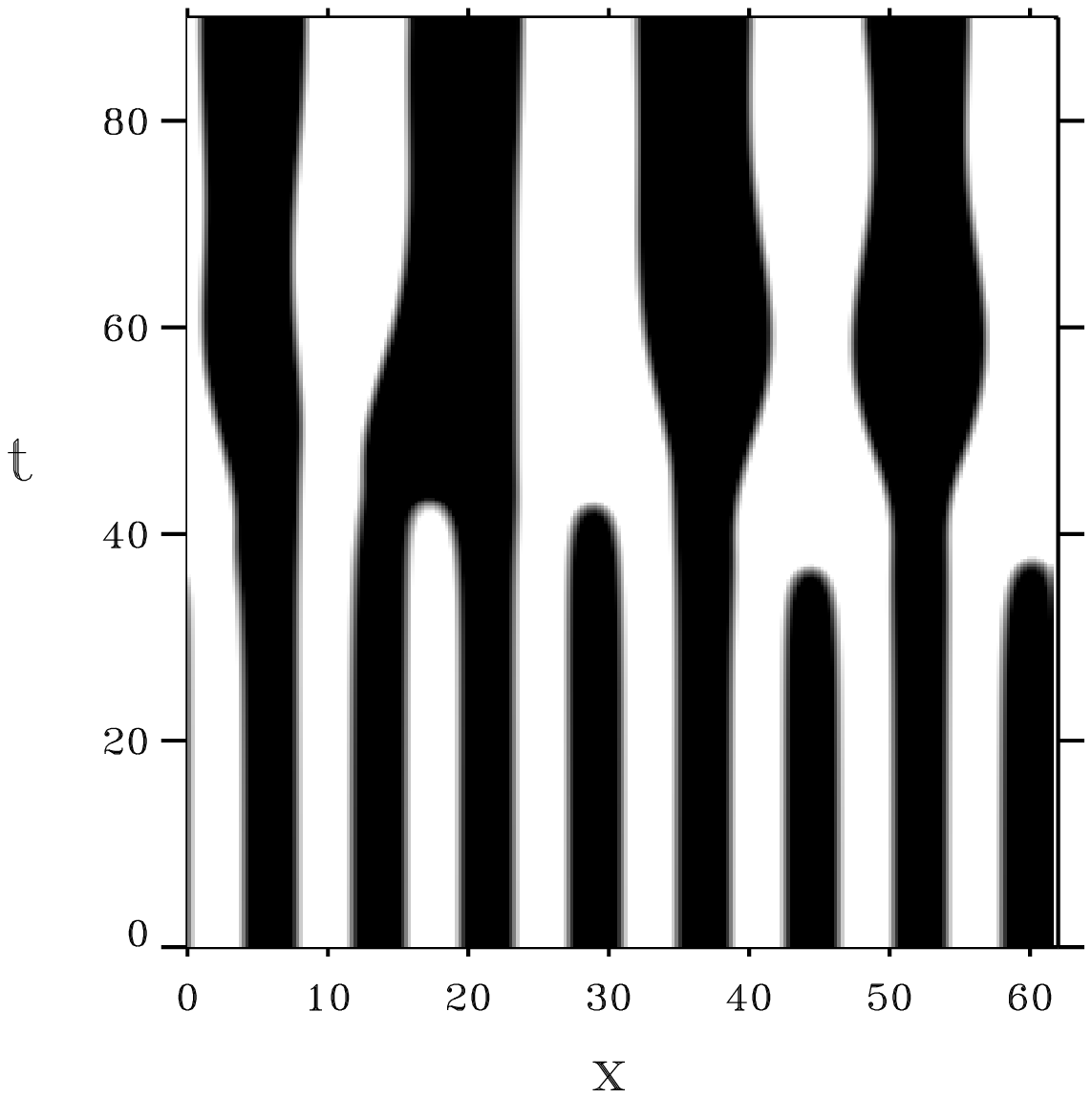}
\caption{
Time evolution of a periodic pattern in the region of Eckhaus
instability.  The high wavenumber pattern is unstable and either
converges to one of the uniform states (top) or a lower wavenumber
pattern (bottom).  Parameters: $a1=2$, $a_0=0$, $\delta=2$; top:
$\epsilon=0.01$, bottom: $\epsilon=0.1$.  }
\label{fig:eckhaus}
\end{figure}

We have also tested the condition for the Eckhaus instability in a
bistable system using numerical computations of $\tau$ and $B$.
Choosing wavenumbers $k>k_c$ where $k_c$ corresponds to the minimum of
$kB$, we found initial periodic patterns either collapse to uniform
states or to a lower wavenumber pattern through phase slips.
Fig.~\ref{fig:eckhaus} demonstrates these two cases.  Similar
conclusions hold for excitable systems.  An unstable Turing pattern,
on the other hand, always converges to a lower wavenumber pattern
since the single uniform state is unstable.

\section{Conclusion}
\label{con}
We have shown that the Cross-Newell phase equation provides a
powerful tool for studying instabilities of stationary periodic patterns in
activator-inhibitor systems. The equation contains information not only
on the Eckhaus and zigzag instabilities, but also on the
destabilization of stationary periodic patterns to traveling waves.
The same equation applies to bistable, excitable, and Turing unstable
systems.  Equations of that kind should prove useful in identifying
parameters and initial conditions where zigzag and Eckhaus
instabilities couple to traveling wave modes.  Such coupling may
lead to complex spatiotemporal behavior analogous to
the coupling of the NIB front bifurcation to a transverse front
instability~\cite{HaMe:94c,HaMe:94b}

\acknowledgments
This study was supported in part by grant  No 95-00112 from the US-Israel
Binational Science Foundation (BSF) and by the Department of Energy, 
under contract W-7405-ENG-36.

%%%%%%%%%%%%%%%%%%%%%%%%%%%%%%%%%%%%%%%%%%%%%%%%%%%%%%%%%%%%%%%
%
%  Appendix
%
%%%%%%%%%%%%%%%%%%%%%%%%%%%%%%%%%%%%%%%%%%%%%%%%%%%%%%%%%%%%%%%
%\clearpage\widetext\onecolumn
\appendix
\section{The Meaning of $\tau=0$}
We show here that the condition $\tau=0$ defines the critical value of
$\epsilon$ at which traveling solutions bifurcate from the stationary
solution.  We look for traveling solutions $u(\theta), v(\theta)$ of
Eqs.~(\ref{fhn}), where $\theta=kx-\omega t$, that bifurcate from the
stationary solution branch $\omega=0$ at some
$\epsilon=\epsilon_c$. Near the bifurcation where $\omega\ll 1$ we can
expand the traveling solutions as power series in $\omega$ around the
stationary solution $u_0,v_0$:
\begin{eqnarray}
u(\theta)=u_0(\theta)+\omega u_1(\theta)+ ... \,, \nonumber \\
v(\theta)=v_0(\theta)+\omega v_1(\theta)+ ... \,,
\label{expansion}
\end{eqnarray}
Expanding $\epsilon$ as
\begin{equation}
\epsilon=\epsilon_c+\epsilon_1\omega +...\,,
\end{equation}
and using these expansions in Eqs.~(\ref{fhn}) we find at order $\omega$
\begin{eqnarray}
\left(k^2\frac{\partial^2}{\partial\theta^2}+1-3u_0^2\right)u_1 - v_1
&=&-{u_0}^{\prime}\,, \nonumber \\ \epsilon_c u_1 +
\left(\delta k^2\frac{\partial^2}{\partial\theta^2} - \epsilon
a_1\right)v_1 &=&-v_0^{\prime}-\epsilon_1(u_0-a_1v_0-a_0)\,.
\label{A_order_1} \nonumber
\end{eqnarray}
Projecting the right hand side  onto
$\left(u_0^{\prime}, -\epsilon_c^{-1} v_0^{\prime}\right)$ gives
\begin{equation}
\epsilon_c=\frac{<{v_0^{\prime}}^2>}{<{u_0^{\prime}}^2>}\,,
\label{epsilon_c}
\end{equation}
where we used Eq.~(\ref{order_0_V}) and switched to the 
notation of a prime for the derivative with
respect to the single argument $\theta$.  Using the
definition~(\ref{tau}) of $\tau$ and~(\ref{epsilon_c}) we find
\begin{equation}
\tau=\left(1-\frac{\epsilon_c}{\epsilon}\right)<{u_0^{\prime}}^2>\,.
\end{equation}
Thus, $\tau=0$ implies $\epsilon=\epsilon_c$ or the onset of traveling
solutions.
%Note that $\epsilon_1$ should be zero because of the symmetry between left
%and right traveling solutions. 

\section{Approximate Stationary Solution} 
For $\mu=\epsilon/\delta\ll 1$ a singular perturbation approach can be
used to approximate the stationary solution
$u_0(\theta),v_0(\theta)$. Rescaling the space coordinate as
$z=\frac{\sqrt{\mu}}{k}\theta$, Eqs.~(\ref{order_0}) become
\begin{eqnarray}
u_0-u_0^3-v_0+ \mu u_0^{\prime\prime} &=&0\,, \nonumber\\
u_0-a_1v_0-a_0+ v_0^{\prime\prime}&=&0 \,, \nonumber 
\end{eqnarray}
where the prime denotes now the derivative with respect to $z$.  Since
the small parameter $\mu$ multiplies the second derivative term
$u_0^{\prime\prime}$, two types of spatial regions can be
distinguished.  Outer regions where $u_0(z)$ varies on a scale of
order unity and the term $\mu u_0^{\prime\prime}$ is negligible, and
inner regions where $u_0(z)$ varies on a very short scale of order
$\sqrt{\mu}$ and the term $\mu u_0^{\prime\prime}$ cannot be
neglected.  In these regions, however, $v_0$ hardly changes.

The analysis of the inner regions leads to the solutions
\begin{equation}
u_0=\pm\tanh{\frac{\theta}{\sqrt{2}k}}, \qquad v_0=0 \,.
\label{inner}
\end{equation} 
These solution represent front structures separating two types of
outer regions: domains of high activator values, $u=u_+(v_0)$ ("up
state"), and domains of low activator values $u=u_-(v_0)$ ("down
state"), where $u_\pm(v_0)$ are the extreme roots of
$u_0-u_0^3-v_0=0$. We look for periodic stationary solutions with
wavelength $\Lambda=\Lambda_- + \Lambda_+$, where $\Lambda_+$ and
$\Lambda_-$ are the widths of up and down states
respectively. Consider now a down state spanning the spatial range
$-\Lambda_-<z<0$ followed by an up state spanning the range
$0<z<\Lambda_+$. The equations for $v$ at these outer regions are
\begin{equation}
v_0^{\prime\prime} - q^2(v_0-v_-)=0, \qquad -\Lambda_-<z<0\,,
\label{down}
\end{equation}
with the boundary conditions $v_0(-\Lambda_-)=v_0(0)=0$, and
\begin{equation}
v_0^{\prime\prime} - q^2(v_0-v_+)=0, \qquad 0<z<\Lambda_+ \,,
\label{up}
\end{equation} 
with the boundary conditions $v_0(0)=v_0(\Lambda_+)=0$. In obtaining
these equations we approximated $u_\pm(v_0)=\pm 1 -v_0/2$. This
approximation is particularly good for bistable media with $a_0$ small
and $a_1$ relatively large. These values restrict $v_0$ to a small
range around $v_0=0$. For excitable media and systems undergoing
Turing instability, a large value of $\delta$ might be needed to keep
$v_0$ small.

The solutions to Eqs.~(\ref{down}) and~(\ref{up}) are
\begin{equation}
v_0=\frac{v_-}{\sinh{q\Lambda_-}}\left[\sinh{qz}-\sinh{q(z+\Lambda_-)}\right]
+v_- \,,
\label{soldown}
\end{equation}
for $-\Lambda_-<z<0$, and
\begin{equation}
v_0=\frac{v_+}{\sinh{q\Lambda_+}}\left[\sinh{q(z-\Lambda_+)}-\sinh{qz}\right]
+v_+ \,,
\label{solup}
\end{equation}
for $0<z<\Lambda_+$. To determine $\Lambda_\pm$ for a given $\Lambda$
we match the derivatives of $v_0$ at the front positions
\[
v_0^{\prime}(0^+)= v_0^{\prime}(0^-), \qquad
v_0^{\prime}(\Lambda_+)=v_0^{\prime}(-\Lambda_-)\,.
\]
This leads to the relation
\[
v_+\beta(q\Lambda_+)+v_-\beta(q\Lambda_-)=0\,,
\]
where $\beta(x)$ is given by Eq.~(\ref{beta}).

\section{Calculation of $\tau$ and $B$}
The quantities $\tau$ and $B$ are given by Eqs.~(\ref{tau})
and~(\ref{B}).  Consider first the integral
\[
<u_0^\prime(\theta)^2>=\frac{1}{2\pi}\int_0^{2\pi}u_0^\prime(\theta)^2d\theta
\,.
\]
It has a contribution from two inner regions at $z=0$ and
$z=\Lambda_+$ where $u_0$ is given by Eq.~(\ref{inner}), and a
contribution from two outer regions, $-\Lambda_-<z<0$ and
$0<z<\Lambda_+$ where $u_0=-1-v_0/2$ and $u_0=1-v_0/2$ with $v_0$
given by Eq.~(\ref{soldown}) and Eq.~(\ref{solup}),
respectively. (Recall that $z=\frac{\sqrt{\mu}}{k}\theta$).

The contribution from the two inner regions is
\begin{eqnarray}
<u_0^\prime(\theta)^2>_{inner}&=&\frac{1}{2\pi k^2}\int_{inner}{\rm
sech}^4 \left(\frac{\theta}{\sqrt{2}k}\right)d\theta \,, \nonumber \\ &
\mbox{} \approx& \frac{1}{\sqrt{2}\pi k}\int_{-\infty}^\infty {\rm
sech}^4 xdx= \frac{2\sqrt{2}}{3\pi k}\,.  \nonumber
\end{eqnarray}
We have used here the fact that $k\sim{\cal O}(\sqrt{\mu})\ll 1$. The
integral over a narrow inner region is transformed into an integral
over a wide region after stretching the $\theta$ variable to the
$x=\frac{\theta}{\sqrt{2}k}$ variable. The contribution from the two
outer regions is
\begin{eqnarray}
<u_0^\prime(\theta)^2>_{outer}&=& \frac{\sqrt{\mu}}{8\pi k} \nonumber
\\ &\times& \left[\int_{-\Lambda_-}^0 {v_0(z)^\prime}^2dz +
\int_{0}^{\Lambda_+} {v_0(z)^\prime}^2dz\right]\,, \nonumber
\end{eqnarray}
where we used in the two outer regions $u_0(z)^\prime=
-\frac{1}{2}v_0(z)^\prime$. Altogether,
\begin{eqnarray}
\label{u_int}
<u_0^\prime(\theta)^2>&=&\frac{2\sqrt{2}}{3\pi k}+
\frac{\sqrt{\mu}}{8\pi k} \\ &\times&\left[\int_{-\Lambda_-}^0
{v_0(z)^\prime}^2dz + \int_{0}^{\Lambda_+}
{v_0(z)^\prime}^2dz\right]\,.  \nonumber
\end{eqnarray}
The second term on the right hand side of Eq.~(\ref{u_int}) is small
(since $\sqrt{\mu}\ll 1$) and will not contribute to the leading order
forms of $\tau$ and $B$.

Consider now the integral
\[
<v_0^\prime(\theta)^2>=\frac{1}{2\pi}\int_0^{2\pi}v_0^\prime(\theta)^2d\theta
\,.
\]
The contribution from the inner regions to this integral is negligible
for $\mu\ll 1$. Thus
\begin{eqnarray}
<v_0^\prime(\theta)^2>&=&\frac{\sqrt{\mu}}{2\pi k}\nonumber\\
&\times&\left[\int_{-\Lambda_-}^0 {v_0(z)^\prime}^2dz +
\int_{0}^{\Lambda_+} {v_0(z)^\prime}^2dz\right]\,.
\label{v_int}
\end{eqnarray}

Using the solutions~(\ref{soldown}) and~(\ref{solup}) in the
integrals~(\ref{u_int}) and~(\ref{v_int}) and using the expressions
for $\tau$ and $B$ we obtain the expressions~(\ref{tauB}).

\bibliography{reaction,hagberg}

\end{document}